\documentclass[conference]{IEEEtran}
\IEEEoverridecommandlockouts
\usepackage{cite}
\usepackage{amsmath,amssymb,amsfonts}
\usepackage{algorithmic}
\usepackage{graphicx}
\usepackage{textcomp}
\usepackage{xcolor}
\usepackage{balance}
\def\BibTeX{{\rm B\kern-.05em{\sc i\kern-.025em b}\kern-.08em
    T\kern-.1667em\lower.7ex\hbox{E}\kern-.125emX}}
\begin{document}

\title{How Helpful do Novice Programmers Find the Feedback of an Automated Repair Tool?}

\author{\IEEEauthorblockN{Oka Kurniawan\IEEEauthorrefmark{1}, Christopher M. Poskitt\IEEEauthorrefmark{2},
Ismam Al Hoque\IEEEauthorrefmark{1},
Norman Tiong Seng Lee\IEEEauthorrefmark{1},\\ Cyrille Jégourel\IEEEauthorrefmark{1}, and Nachamma Sockalingam\IEEEauthorrefmark{1}}
\IEEEauthorblockA{\IEEEauthorrefmark{1}Singapore University of Technology and Design, Singapore}
\IEEEauthorblockA{\IEEEauthorrefmark{2}Singapore Management University, Singapore}}

\maketitle

\begin{abstract}
    Immediate feedback has been shown to improve student learning.
    In programming courses, immediate, automated feedback is typically provided in the form of pre-defined test cases run by a submission platform.
    While these are excellent for highlighting the presence of logical errors, they do not provide novice programmers enough scaffolding to help them identify where an error is or how to fix it.
    To address this, several tools have been developed that provide richer feedback in the form of program repairs.
    Studies of such tools, however, tend to focus more on whether correct repairs can be generated, rather than how novices are using them.
    In this paper, we describe our experience of using CLARA, an automated repair tool, to provide feedback to novices.
    First, we extended CLARA to support a larger subset of the Python language, before integrating it with the Jupyter Notebooks used for our programming exercises.
    Second, we devised a preliminary study in which students tackled programming problems with and without support of the tool using the `think aloud' protocol.
    We found that novices often struggled to understand the proposed repairs, echoing the well-known challenge to understand compiler/interpreter messages.
    Furthermore, we found that students valued being told where a fix was needed---without necessarily the fix itself---suggesting that `less may be more' from a pedagogical perspective.
\end{abstract}


\section{Introduction}
One of the most common methods of assessing students in computing is practical work, where students are tasked to solve programming problems~\cite{Carter2003}.
The large enrollments in computer science courses, however, make it more challenging for faculty to provide one-to-one feedback to students on their programming assignments.
During such practical work, novice programmers often encounter feedback on their solutions through compiler or interpreter error messages, such as those arising from syntax errors, and information from automated assessment tools, such as the number of test cases passed or failed~\cite{Ihantola2010}.
While proficient programmers are able to make use of test cases to identify some logical flaws or boundary case bugs in their programs, novice programmers often find it difficult to do so~\cite{Ettles2018}.
There is a need, then, for novice programmers to be guided through their logical errors and on how to fix them.
Are we able to automatically and accurately generate this feedback for novice programmers?

One definition of feedback is ``information communicated to the learner that is intended to modify his or her thinking or behaviour for the purpose of improving learning''~\cite{Shute2008}. For error messages, as well as messages from automated assessment tools, Keuning et al.~classifies such feedback under the category of ``Knowledge about Mistakes''~\cite{Keuning2018}. However, in spite of the fact that information contained in error messages can lead programmers to a correct solution, it is known that novice programmers find error messages challenging to interpret and apply to their work~\cite{Becker2019}.  
Keuning et al.~describes another category of feedback as ``Knowledge on how to proceed'', which provides learners information on how to get towards a solution~\cite{Keuning2018}. Paiva et al.~describes three types of tools that provide such feedback, the first of which recommends ``a possible correction when the learner encounters a bug''~\cite{Paiva2022}. Automated program repair~(APR) tools, which can provide hints to novice programmers on how their solution can be corrected, fall under this category. Examples of such tools include CLARA~\cite{Gulwani2018}, Refactory~\cite{Hu2019}, AssignmentMentor~\cite{Li2022}, and Verifix~\cite{Ahmed2022}. While the effectiveness of these APR tools at constructing patches has been experimentally assessed, less attention has been given to how these fix suggestions are used by novice programmers in their learning.

In this paper, we report on our experience of using one such APR tool, CLARA~\cite{Gulwani2018}, to support the teaching of programming to novices.
First, we extended the original software to `CLARA-S' (for CLARA Service), which covers more of the Python language (e.g.~import statements, lambda functions, object-orientation) and exposes CLARA's functionality over a REST API.
Next, we integrated CLARA-S into the Jupyter Notebooks used by our students in their programming exercises, allowing them to request a fix by clicking a button.
Finally, we designed a preliminary study to explore how students were using the feedback generated by CLARA-S.

\section{Related Work}

Singh et al.~proposed an APR tool for introductory programming assignments~\cite{Singh2013}.
This feedback generator requires instructors to specify the correct solution together with a list of rules to apply when correcting errors.
The APR tool of Parihar et al.~\cite{Parihar2017} also uses rules targeted at various compile-time errors in C.
Others have used machine learning techniques on existing student solutions (e.g.~from previous course runs).
For example, clustering can be used to group correct solutions as input for repair programs~\cite{Glassman2015,Gulwani2018}.
Such tools find the `nearest' cluster of correct solutions and propose a repair based on it.
CLARA, the APR tool used in this report, applies the clustering approach to C and Python programs, and has achieved a repair rate of 97.44\% over four thousand incorrect Python programs~\cite{Gulwani2018}.

Developing APR tools continues to be an active area of research with a diversity of approaches~\cite{Paiva2022}.
The technical approach taken by Refactory is to refactor all student solutions to match the structure of a single correct solution~\cite{Hu2019}.
Other tools include AP-Coach~\cite{Duong-Shar-Shankararaman22a}, AssignmentMender~\cite{Li2022} and Verifix~\cite{Ahmed2022}, all of which use different technical approaches.
AP-Coach, for example, uses code similarity techniques to find the closest reference solution, then generates feedback in the form of comments and AI-generated pseudocode.

Some preliminary user studies have been described by authors of APR tools.
For CLARA~\cite{Gulwani2018}, the average rating of 52 participants was 3.4 out of 5, with 5 indicating that the feedback provided was ``most useful''.
While textual feedback was collected, it was not provided by the study.
It was reported that more students found the program repair feedback generated by AssignmentMender helpful compared to the feedback generated by Refactory~\cite{Li2022}.
A survey was designed showing instances of an incorrect program and the repair provided by Verifix~\cite{Ahmed2022}.
This was administered to 14 tutors of introductory programming courses who gave generally positive feedback on the tool.
While encouraging, these studies give limited guidance to practitioners who may wish to use APR tools.

Two studies attempted to understand the benefit of APR tools for novice programmers~\cite{Ahmed2022, Reis2019}.
Ahmed et al.~\cite{Ahmed2022} used a data-driven approach to compare the use of automated feedback tools to human tutors in two cohorts of students, and concluded that the benefit of automated feedback tools was primarily to provide an alternative source of feedback in the absence of a human tutor.
However, no pedagogical benefit was found: under exam conditions, students in the automated tools cohort were not seen to fix errors faster than the students in the human tutors cohort.
Reis et al.~\cite{Reis2019} assigned 42 novices into three groups, tasking them to solve programming problems.
All three groups had access to test cases, with the second having access to CLARA, and the third having access to PythonTutor (a web-based visualiser of Python program executions~\cite{Guo2021}).
Students who had access to CLARA solved the problems faster and rated the feedback as more useful, but there was no difference in performance in a programming post-test between the CLARA group and test cases (only) group.

\section{Context}

\emph{\textbf{Our Courses}.}
Our intervention took place at the Singapore~University~of~Technology~and~Design, which requires all students to take three terms of common courses.
This includes two courses on programming, both taught using Python.
The first course is compulsory for all students in their first term (Term 1), while the other is an elective taken in their third term (Term 3), and is strongly recommended for those choosing to major in computer science.

The Term 1 course focuses on basic syntax, control structures, and some object-oriented programming.
The Term 3 course covers basic sorting algorithms, data structures, and further topics on object-oriented programming.
(It can thus be assumed that students who completed Term 3 are generally more proficient at programming.)
Both courses utilise Jupyter Notebook for their programming assignments.

\emph{\textbf{CLARA-S}.}
The original version of CLARA~\cite{Gulwani2018} supports only a subset of Python and must be run in a Linux terminal.
We modified its parser so that it could additionally support import statements, lambda functions, and some object-oriented programming features.
In particular, this enables CLARA to parse programming assignments that import and utilise built-in functions from other libraries.

\begin{figure}[!t]
\includegraphics[width=\linewidth]{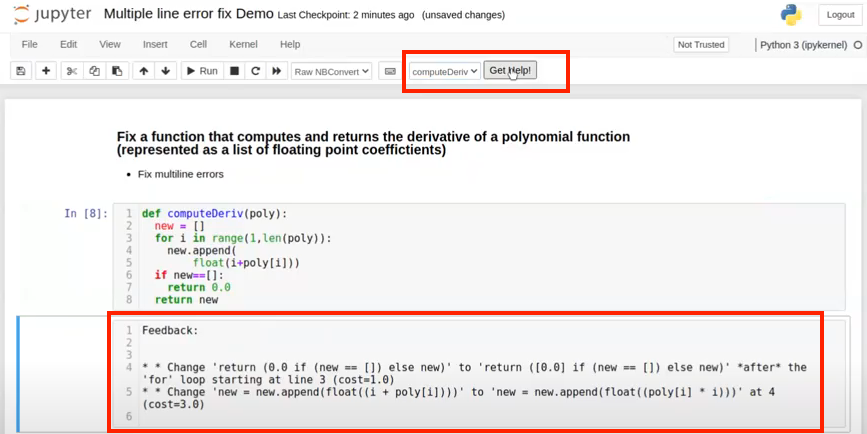}
\caption{Screenshot of our Jupyter extension to call CLARA}
\centering
\label{fig:extension}
\end{figure}

Using FastAPI, we exposed CLARA as a RESTful web service~\cite{Al_Hoque_API_Server_for_anonymous}, allowing it to be invoked over HTTP requests from any other application.
Furthermore, we developed a Jupyter Notebook extension~\cite{Al_Hoque_Jupyter_Notebook_Extension_anonymous}, allowing students to call the CLARA service using a button in our programming exercises.
We refer to the overall solution as `CLARA-S' (for CLARA Service); the source code is available online~\cite{Al_Hoque_CLARA-S_anonymous}.
A screenshot of our extension is shown in Fig.~\ref{fig:extension}.
At the top, we highlight the button that students press to trigger feedback while working on a programming exercise.
At the bottom, we highlight some feedback from CLARA-S, which the extension automatically inserts into a new cell of the Jupyter Notebook.

CLARA-S requires a repository of correct solutions~\cite{Gulwani2018}, which its underlying mechanism then reduces into a cluster representation (each cluster intuitively representing a different solution approach).
For instructors, one way to populate this repository is to utilise the (correct) solutions submitted to programming assignments in previous runs of the course.
Repair feedback is then generated based on the minimum distance to one of those cluster representations.
From a student's point of view, they access CLARA-S directly from their Jupyter notebook---in particular, an online platform where the notebook has been set up containing questions in markdown, as well as starter code and test cases in some code cells.
Students are then able to generate feedback for their incorrect solutions by choosing the function name and clicking the integrated help button in the interface (Fig.~\ref{fig:extension}), after which the feedback appears in a new cell.

\section{Study Protocol}
\label{sec:study_protocol}

\begin{figure}[!t]
\includegraphics[width=\linewidth]{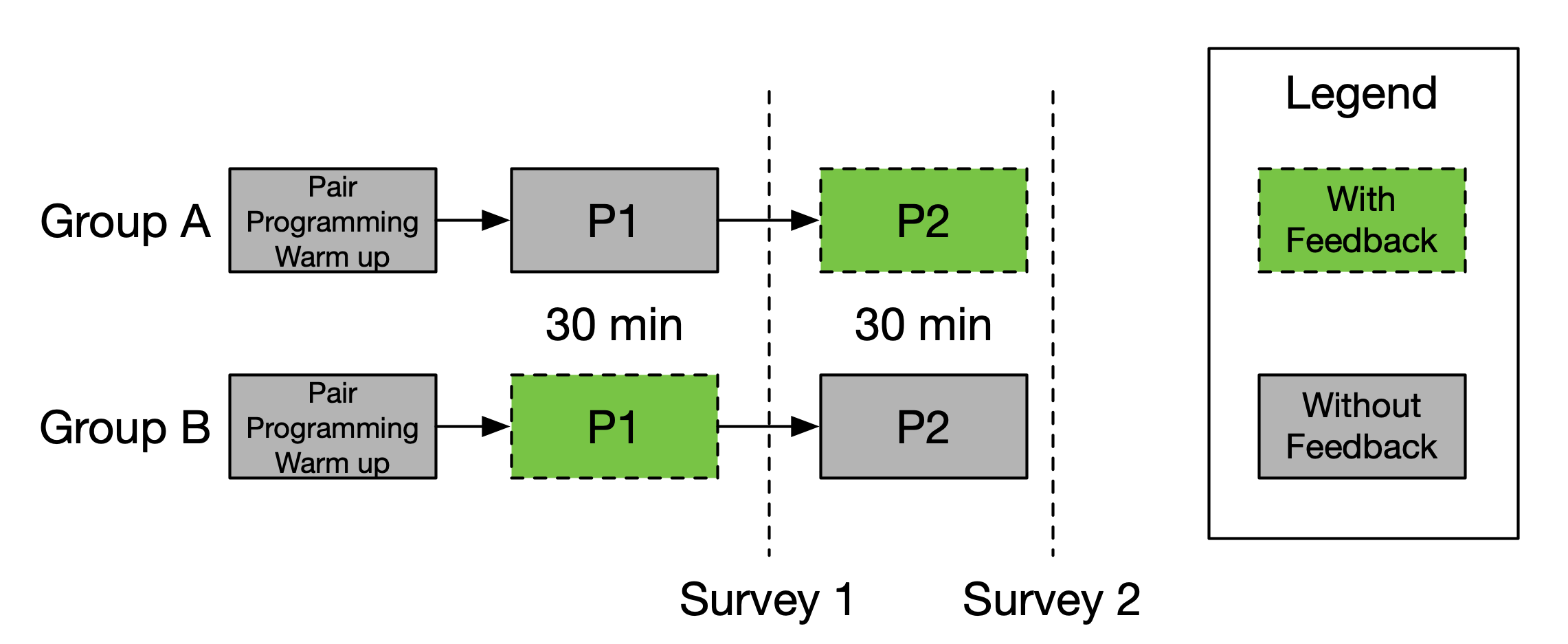}
\caption{Study protocol for the two groups of students, Group A and B. Each participant is given 30 mins for each programming task, P1 and P2. Group A uses CLARA-S for P2, whereas Group B uses it for P1}
\centering
\label{fig:protocol}
\end{figure}

We designed a preliminary study to gain some insights into how helpful our students find the feedback provided by CLARA-S, and to help us understand where further research and tool improvements should be focused.
Our study protocol is shown in Fig.~\ref{fig:protocol}.
First, the participants were randomly split into two groups (A and B).
In each group, participants started with a warm-up exercise before being tasked to solve two programming problems in sequence: P1 and P2.
Participants in Group A first attempted P1 without being allowed to use CLARA-S, followed by P2 with CLARA-S.
In Group B, this sequence is reversed: participants attempted P1 with CLARA-S followed by P2 without.
Participants were not forced to use CLARA-S when it was allowed.

P1 and P2 were chosen to be of novice and intermediate difficulty respectively: P1 required implementing a function that finds the position of an item in a sorted array, whereas P2 required implementing base 10 addition for numbers represented as linked lists of digits.
As CLARA-S needs to be provided some correct solutions for clustering, the research team independently solved the exercises and produced several different correct solutions.
CLARA-S reduced these correct solutions to three clusters for task P1 and four for task P2.

The participants were to attempt P1 and P2 using pair-programming together with a research assistant.
Participants were instructed (through a warm-up exercise) to use a think-aloud protocol, i.e.~to verbalise their thoughts in solving the problem to the research assistant, who would then write the code.
The research assistant was not allowed to provide any help if the participants experienced difficulties regarding logical or semantic errors, but was allowed to clarify language syntax.
This was to isolate participants' difficulties when using CLARA-S so that feedback is only given for logical or semantic errors (our focus).
The participants were given a maximum of 30 minutes per task and could stop at any time.

At the end of each programming problem, the participants were asked to fill in an online survey.
The survey was based on \cite{Kolar2013} and \cite{Gok2012}.
The first set of questions was used to measure the computing self-efficacy of the participants.
The second set of questions was used to measure problem solving confidence using a validated instrument designed by Gok~\cite{Gok2012}.
As Gok's instrument was originally designed for physics problem-solving, we adapted the wording of the questions to suit the programming problem-solving context.
In addition, if participants had just solved P1 or P2 with the help of CLARA-S, the survey contained additional questions on their experience and perception of using it. 

In order for the participants to get used to the think-aloud protocol~\cite{Fan2020} and pair programming, both groups of participants began with a warm-up exercise.
In this warm-up exercise, the participants were given three simple programming tasks.
For the first task, the research assistant demonstrated the think-aloud and pair programming protocols.
For the next two tasks, the participants solved the problems by trying these protocols themselves.
The outcome of the warm-up exercises were used by the research assistant to identify the programming background of the participants.
The participants were categorised as `novice', `intermediate', or `advanced' according to how they solved the warm-up exercises.

We invited students taking our Term 1 and Term 3 programming courses (in our academic calendar, these terms coincide) to take part in our study.
Participation was explicitly voluntarily.
We obtained 20 participants, who were grouped randomly into Group A and Group B (10 in each).
Group A consisted of 5 novice and 5 intermediate programmers, whereas Group B consisted of 3 and 7 respectively.

\section{Results and Discussion}

\emph{\textbf{Usage of CLARA-S.}}
We observed in our preliminary study that when given the option, seven participants from Group A and five from Group B chose to use CLARA-S.
The largest drop in the latter came from intermediate programmers: only two of the seven intermediate programmers used it, with the remainder able to complete the problem without hints.
A likely explanation is that task was simple enough for those programmers to solve without needing feedback.
Another factor could be due to the nature of the study: participants may have been driven to demonstrate that they were strong enough to solve the problems without any assistance.

\begin{table}[!t]
    \centering\footnotesize
        \caption{Normalised Likert scores with respect to background}    \label{tab:background}
\begin{tabular}{|lrr|}\hline
  Question                                      & Novice                    & Intermediate              \\\hline
The programming tasks are difficult       & 0.19                      & -1.06                     \\
I have enough time to complete the tasks & -0.25                     & 1.35 \\\hline                     
\end{tabular}
\end{table}

Table \ref{tab:background} summarises the survey questions on task difficulty according to background.
We use normalised Likert scores, where -2.0 (resp.~+2.0) indicates universal strong disgreement (resp.~agreement).
Novice programmers slightly agreed that the tasks were difficult, whereas the intermediate programmers found them easier (explaining why fewer required fixes).
Similarly, while novices felt they did not have enough time, intermediate programmers tended to agree that they did.
We also measured the time taken by participants.
The average duration for P1 was 18.05 minutes; for P2, it was 24.50 minutes.
This suggests that both tasks were of moderate difficulty, with P2 requiring more thought (as expected).

\begin{figure*}[!t]
\includegraphics[width=\linewidth]{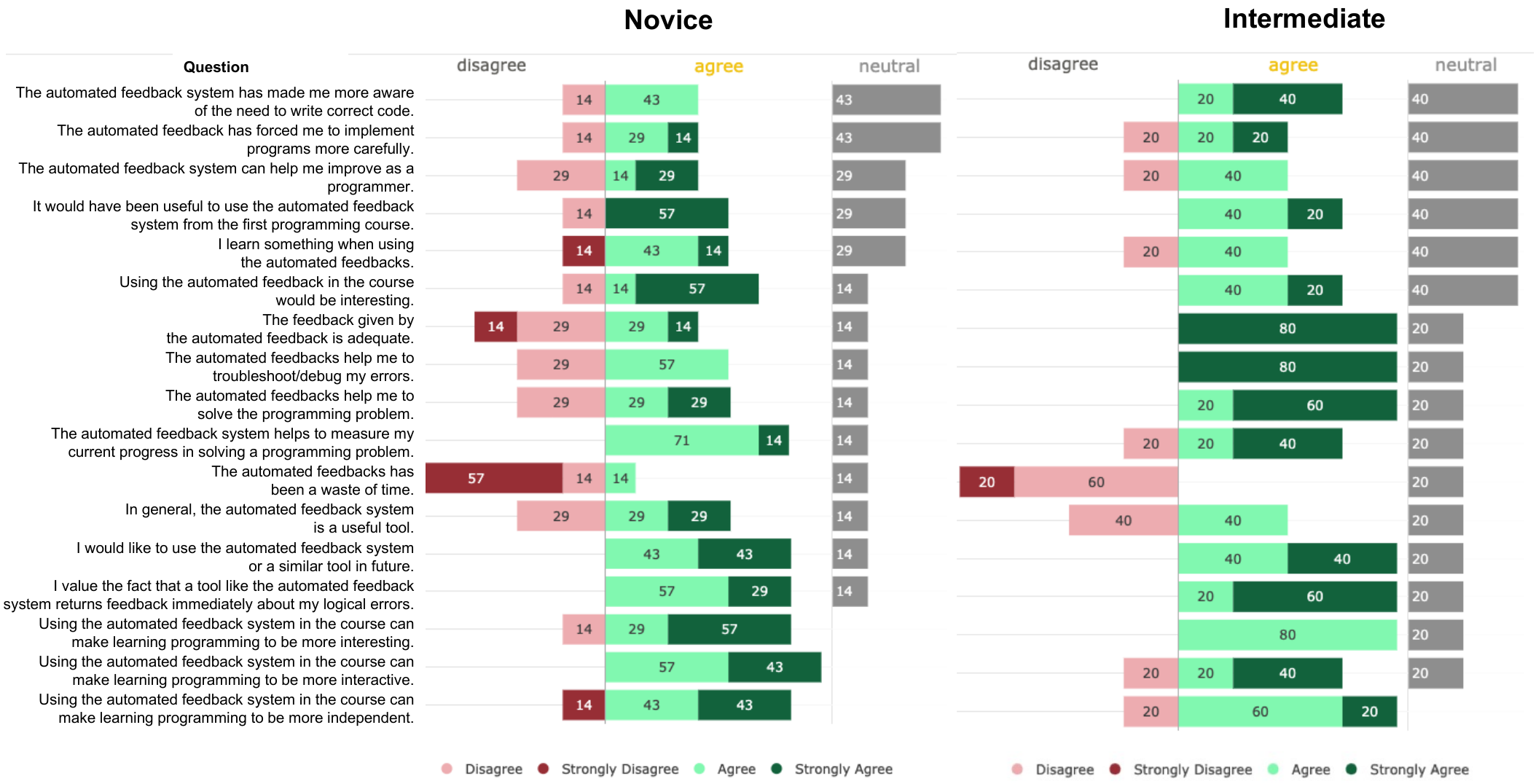}
\caption{Feedback on the use of CLARA-S with respect to programming background. The numbers in the bars are percentages}
\centering
\label{fig:feedback_likert_background}
\end{figure*}

\emph{\textbf{Helpfulness of CLARA-S.}}
Fig.~\ref{fig:feedback_likert_background} summarises the survey results, categorised by background, regarding the usefulness of the feedback (i.e.~repairs) suggested by CLARA-S.
It can be seen that participants with intermediate programming backgrounds tended to give more positive feedback on the tool compared to novice programmers: while novice programmers were divided 50-50 on whether the feedback given is adequate, the intermediate programmers were more positive on this aspect.
Though both groups disagree that the APR tool is a waste of time, we can see that a higher percentage of novice programmers chose `strongly disagree'.
This means amidst its drawbacks, the novice programmers still perceive that CLARA-S helps them in some way.
When categorised by Group A vs.~B, the results show that 86\% of Group A---who were allowed to use CLARA-S for the more challenging task, P2---found the tool useful compared to just 14\% of Group B.

\emph{\textbf{Deciphering the Feedback.}}
Three of the novice programmers reported that they found it difficult to decipher the feedback messages given by CLARA-S.
We followed-up with them in an informal interview, receiving the following comment:

\begin{quote}\small
    \emph{``When I got an error, I tried to use [CLARA-S]. I did not understand the feedback.''}
\end{quote}

\begin{figure}[!t]
\centering
\includegraphics[width=0.5\linewidth]{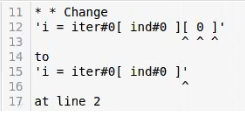}
\caption{Example of CLARA-S output seen by a participant}
\label{fig:clara_output}
\end{figure}

An example of CLARA-S output is shown in Fig. \ref{fig:clara_output}. The caret symbol in the output is an additional feature we added on top of CLARA to show which parts of the line to repair.
As can be seen, even though CLARA specifies the repair to be done, the message can be cryptic for novice programmers.
In particular, CLARA gives its feedback via an internal representation (e.g.~$\mathtt{iter\#0}$ and $\mathtt{ind\#0}$) rather than using exactly the code written by the student.
The problem is analogous to the issues novices face interpreting compiler/interpreter message for syntax errors.
Several studies explore how to make these error messages more novice-friendly (e.g.~\cite{NienaltowskiPM08}), and we believe that the results of these studies as well as resugaring techniques~\cite{Pombrio2015} may be useful to incorporate into APR tools that help novices fix semantic/logical errors.

Most participants only used the feedback system when they got stuck, while a few participants used it to check their logic and answers.
Most students used it only when they could not proceed, as shown in the following comments:

\begin{quote}\small
    \emph{``When I could not solve the questions anymore, I started using the automated system.''}
    
    \emph{``I made use of it when I felt like I was stuck and needed more help.''}
\end{quote}

Some used it immediately upon a failing test case:

\begin{quote}\small
    \emph{``If my code is running into an error, or not passing the asserts. I would press feedback and see the suggestions.''}
\end{quote}

One participant made use of CLARA-S to guide his implementation.
He knew he required a loop, so started writing the structure of the loop and fed it into CLARA-S to get feedback on the detail of the for-loop code: 

\begin{quote}\small
    \emph{``I created a simple for loop function (that would mimic what is expected of the final function) in order to get the automated system to feed me answers and use the feedback until correct solution is reached.''}
\end{quote}

\emph{\textbf{Location vs.~Fix.}}
An interesting finding from the study is that students found the current version of CLARA-S most helpful for pointing out \emph{where} the logical error is, rather than \emph{what} the solution should be.

\begin{quote}\small
    \emph{``It is good that there are arrows pointing to which part I should change. But the alternative solution was very different from mine, it was quite confusing.''}
\end{quote}

The participant above found the feedback to be helpful for locating which part they needed to change, but the proposed repair caused confusion.
This is especially true for novice programmers and can be seen by the ambivalent result of the survey on whether the feedback is adequate (Fig.~\ref{fig:feedback_likert_background}).
This could also due to the limited number of correct solutions fed into the system: better feedback may be provided with additional cluster representations.

\begin{quote}\small
    \emph{``It's a bit confusing, possibly due to first time usage. Hunch is that if it's closer to the solution it's easier~to~use.''}
    
    \emph{``I think it is good to guide students what to do. But I feel that it restricts to one certain solution only.''}
\end{quote}

\emph{\textbf{Reflections.}}
So what can we conclude from our experience of using CLARA-S?
First, we have observed that automated repair suggestions provide useful feedback for students with different backgrounds, especially for more complex tasks.
The feedback message seems to benefit intermediate programmers more than the novice programmers, likely due to the cryptic nature of some feedback generated by the system, echoing the similar challenges novices face with compiler messages.
Nonetheless, students valued being shown \emph{where} a fix needed to be applied, i.e.~the source of the logical error.

Another benefit that we see from this system is for instructors to speedily localise logical errors.
When grading buggy solutions or consulting with students, instructors can use CLARA-S to immediately hone in on the critical part of the code.
In other words, the tool's feedback can still be used effectively, even in instances where the student cannot decipher it independently.
This aligns with Yi~et~al.~\cite{Yi-et_al17a}, who observed that repairs helped teaching assistants grade more efficiently.

Finally, from a pedagogical perspective, we would like students to be able to find the solutions themselves without just giving them a complete fix (as CLARA-S currently does).
An incremental APR-based hinting system may be better for developing students' problem solving skills and higher order thinking.
For example, the first `hint level' could simply localise the logical error, i.e.~the tool informs students which line needs fixing (without suggesting how).
From here, further hint levels could gradually reveal details of the fix until the whole repair is given (akin to the hint system of~\cite{AntonucciENPM15}).

\section{Conclusion}

We developed CLARA-S, an extension of the CLARA APR tool that covers more of the Python language and can be run as a Jupyter Notebook extension.
In order to understand how students perceive it as an aid for learning programming, we designed a preliminary study in which students tackled exercises with and without support from the tool using a think aloud protocol.
While students expressed positive sentiments overall, we found that novices struggled to decipher the feedback that CLARA-S gave to them.
This was reminiscent of the known challenges they face with compiler messages, suggesting that usability work in that area might be applicable here too.
We also found that intermediate programmers and instructors benefited most from being told where the fix needed to be applied (more so than how they needed to fix it).

Our immediate task is to expand this preliminary work into a full study and measure the impact on learning with a larger group.
We also want to explore how the feedback from CLARA-S can be displayed/explained in a way that better supports novices.
Furthermore, future work should explore the potential utility of an APR-driven incremental hint system akin to~\cite{AntonucciENPM15}, i.e.~by initially revealing only the fix location before gradually revealing (partial) repairs.
It would also be interesting to study how hint-based APR systems impact learning in comparison to systems that immediately reveal the complete solution, such as ChatGPT~\cite{chatgpt}.

\section*{Acknowledgements}
This research was funded by SUTD under the Pedagogy Leadership Grant (PLG) scheme.
We are grateful to Jasmine~Tan (SUTD-IRB) for her helpful feedback and suggestions on our study protocol.

\bibliographystyle{IEEEtran}
\balance
\bibliography{references}

\end{document}